\documentclass[aps,pra,twocolumn,showpacs,superscriptaddress]{revtex4-1} 
\usepackage{graphicx,amssymb,amsfonts,amsmath}

\usepackage{ulem}

\newcommand{\etal}{ \text{\it et al.} }
\newcommand{\CO}{(Color online)\;}

\newcommand{\be}{\begin{eqnarray}}
\newcommand{\ee}{\end{eqnarray}}

\newcommand{\n}{\nonumber \\}
\newcommand{\piboost}{$\pi-$boost }

\usepackage[utf8]{inputenc} 
\usepackage{graphicx}
\usepackage{bbold}

\usepackage[colorlinks,bookmarks=false,citecolor=blue,linkcolor=red,urlcolor=blue]{hyperref}

\begin{document}

\title{Relaxation towards negative temperatures in bosonic systems: Generalized Gibbs ensembles and beyond integrability}
\author{Stephan Mandt}
\affiliation{Princeton Center for Complex Materials and Department of Physics, Princeton University, New Jersey 08544, USA.}
\author{Adrian E. Feiguin}
\affiliation{Department of Physics, Northeastern University, Boston, Massachusetts 02115, USA.}
\author{Salvatore R. Manmana}
\affiliation{Institute for Theoretical Physics, University of G\"ottingen, D-37077 G\"ottingen, Germany.}
\date{\today}

\begin{abstract}
Motivated by the recent experimental observation of negative absolute temperature states in systems of ultracold atomic gases in optical lattices [Braun {\it et al.}, Science {\bf 339}, 52 (2013)], we investigate theoretically the formation of these states. 
More specifically, we consider the relaxation after a sudden inversion of the external parabolic confining potential in the one-dimensional inhomogeneous Bose-Hubbard model.
First,
we focus on the integrable hard-core boson limit which allows us to treat large systems and arbitrarily long times, 
providing convincing numerical evidence for relaxation to a generalized Gibbs ensemble at negative temperature $T<0$,
a notion we define in this context. 
Second, going beyond one dimension, we demonstrate that the emergence of negative temperature states can be understood in a dual way in terms of positive temperatures, 
which relies on a dynamic symmetry of the Hubbard model. 
We complement the study by exact diagonalization simulations at finite values of the on-site interaction. 
\end{abstract}


\maketitle

\section{Introduction}

Since their first realization in nuclear spin systems in the 1950s~\cite{purcell, ramsey}, negative absolute temperatures have become canonical in physics education~\cite{dyson, llifschitz, kittel}. 
Misleadingly, they do not refer to thermal states below absolute zero, but instead describe systems where high energy states are more likely to be occupied than low-energy states. 
Just as systems at positive temperature need an energy minimum, i.e. a ground state, in order to 'pile up' in low energy states~\cite{carr}, negative temperatures need an energy {\it maximum}. 
This upper bound in the energy spectrum and a good insulation from the environment (which is usually at $T>0$) are the necessary ingredients for the realization of negative temperatures, see Refs.~\onlinecite{ramsey, carr, braun} for more details. 
Notably, all laws of thermodynamics apply to systems at negative temperatures, which makes them distinct from the more general class of systems in nonequilibrium steady-states with inverted energy populations, such as lasers. 

Ultracold atoms in optical lattices provide one of the best controlled experimental setups to explore quantum many-body physics. 
In contrast to materials or mesoscopic devices, they are free of phonons, impurities or lattice defects, and hence can serve as quantum simulators of condensed matter model Hamiltonians (for a review see, e.g., Ref.~\onlinecite{blochReview}). 
Since atoms in optical lattices move much slower than electrons in materials, they have proven to be ideal for the study of 
out-of-equilibrium many-body dynamics, see e.g. Refs.~\onlinecite{blochCollapse,newtonsCradle,naegerl1, exp1Dcloud, naegerl2, expansion}. 
 \begin{figure}
 \begin{center}
  \includegraphics[width=\linewidth]{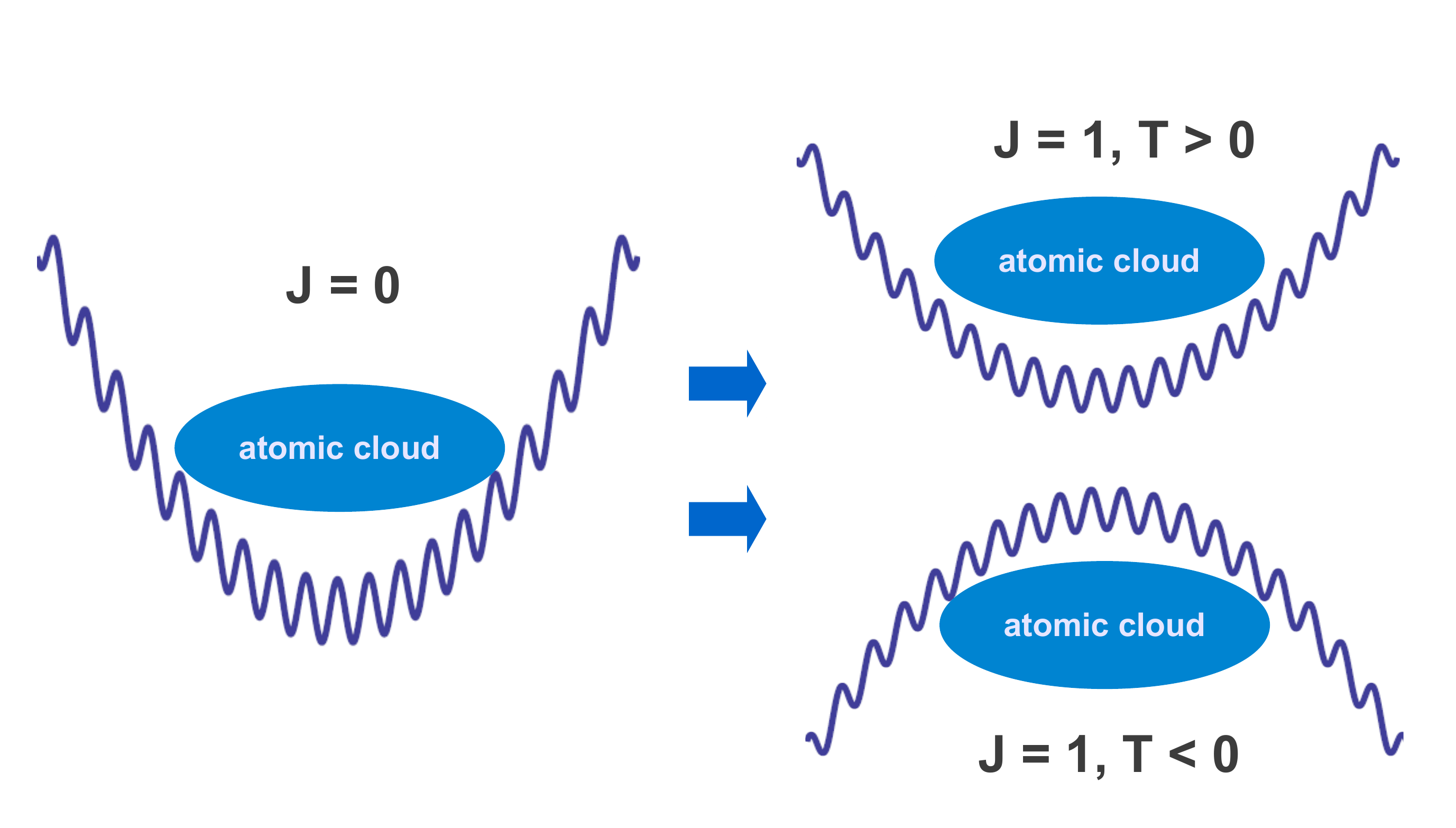}
 \caption{~\CO Graphical illustration of the quench dynamics studied in this paper. Starting from the ground state of the system at $J=0$ (a product state), we study the quenches
	  towards $J=1$ combined with either $V\rightarrow V$ or $V\rightarrow -V$. Both quenches are dual to each other: while $V>0$ results in a final state at $T>0$ (top), negative absolute temperatures emerge for $V<0$ (bottom).}
 \label{fig:sketch}
\end{center}
\end{figure}
Very recently, Braun\etal~\cite{braun} for the first time have realized negative absolute temperatures in a system with motional degrees of freedom, as given in a gas of ultracold bosonic atoms in an optical lattice. 
In a nutshell, the experiment made use of the fact that the kinetic energy in lattice systems is bounded not only from below but also from above, and hence interactions and the trapping potential determine the boundedness of the total energy spectrum. 
In order to reach negative temperatures, several system parameters are ramped in real time, as proposed in earlier theoretical work~\cite{mosk, negT}: 
first, an atomic cloud is prepared in a deep optical lattice, so that tunneling is suppressed and the atomic motion is completely frozen. 
Second, the external potential is inverted, i.e. the trap is turned into an anti-trap, and third, the lattice depth is reduced again so that the atoms are free to move, see Fig~\ref{fig:sketch}. 
It has been argued earlier~\cite{negT} that even when ramping the system parameters infinitely slowly (impossible in experiment), the relevant equilibration times will also diverge, and hence the adiabatic limit is out of reach. 
Therefore, the realization of negative temperatures naturally involves nonequilibrium dynamics. 
Also conversely, negative temperatures may often emerge in simple nonequilibrium setups with cold atoms, such as interacting clouds in tilted lattices~\cite{gravity}.
From an applied perspective as proposed in Ref.~\onlinecite{akosQS}, negative temperatures might help to realize effective attractive interactions for atoms such as ${}^{173}{\rm Yb}$ where formerly only repulsive interactions had been accessible experimentally, and hence may help to realize novel phases of matter.  

The process of equilibration towards negative temperature states is the focus of this paper. 
In contrast to earlier works on fermionic Boltzmann-dynamics~\cite{negT} and nonequilibrium mean field theory~\cite{akosNEQ}, the full quantum dynamics is taken into account. 
We idealize the process in terms of a simultaneous \textit{quench} both in the external potential and in the hopping rate. 
Besides simulating the process numerically for the one-dimensional Bose-Hubbard model for infinite and finite on-site interaction strength, we also provide a picture of the relaxation process in \textit{any} dimension: generalizing a dynamic symmetry of the Hubbard model~\cite{expansion}, we can map the process to a dual relaxation process that ends up at positive temperatures. 

The paper is organized as follows. In Sec.~\ref{sec:model}, we introduce the model and generalize the dynamic symmetry found in Ref.~\onlinecite{expansion} to explain
the duality in the equilibration towards positive and negative temperatures. In Sec.~\ref{sec:hcb}, we summarize the numerical approach used
and discuss our numerical data of the quenched system. We compare the long-time averaged data with reference equilibrium ensembles 
in Sec.~\ref{sec:GGE}, where we coin the notion of negative-temperature generalized Gibbs ensembles.
In Sec.~\ref{sec:ED}, we present exact diagonalization results for the nonintegrable case, before we summarize our results in Sec.~\ref{sec:summary}.

\section{Model and dynamic symmetry}
\label{sec:model}
The focus of this paper is the inhomogeneous Bose-Hubbard Model, which is a good approximation to ultracold bosonic atoms in optical lattices~\cite{blochReview}:
\be
\label{Hubbard}
H (J,U,V) & =&  -J\sum_{<i,j>}(b_i^\dagger b_j + {\rm H.c.}) \\
	  & & + \frac{U}{2}\sum_i n_i(n_i-1)  + V \sum_i x_i^2 n_i \nonumber
\ee 
The first term describes the hopping of the bosonic atoms between nearest neighboring lattice sites, while the second term
takes local on-site interactions into account. The third term models an additional (anti-)confining potential, depending on the sign of $V$. Note that the last term 
explicitly breaks translational invariance.
As mentioned earlier, it is important to realize that this Hamiltonian is bounded from below for $(U>0,V>0)$, while it becomes bounded from above for $(U<0,V<0)$.
The Hamiltonian is unbounded both for $(U<0,V>0)$ and $(U>0,V<0)$ and hence equilibration is prevented in those cases.

We idealize the experimental protocol to reach negative absolute temperatures in the following way:
\begin{itemize}
 \item[1.] The system is initially prepared in the ground state for $J=0$, $U>0$ and $V>0$. 
 \item[2.] The trapping potential and the interaction strength are inverted, i.e. $V\rightarrow -V$ and $U\rightarrow -U$.
 \item[3.] Simultaneously, the hopping rate is switched to $J=1$, and the system is let to evolve in time.
\end{itemize}

\begin{figure}
\begin{center}
\includegraphics[width=\linewidth]{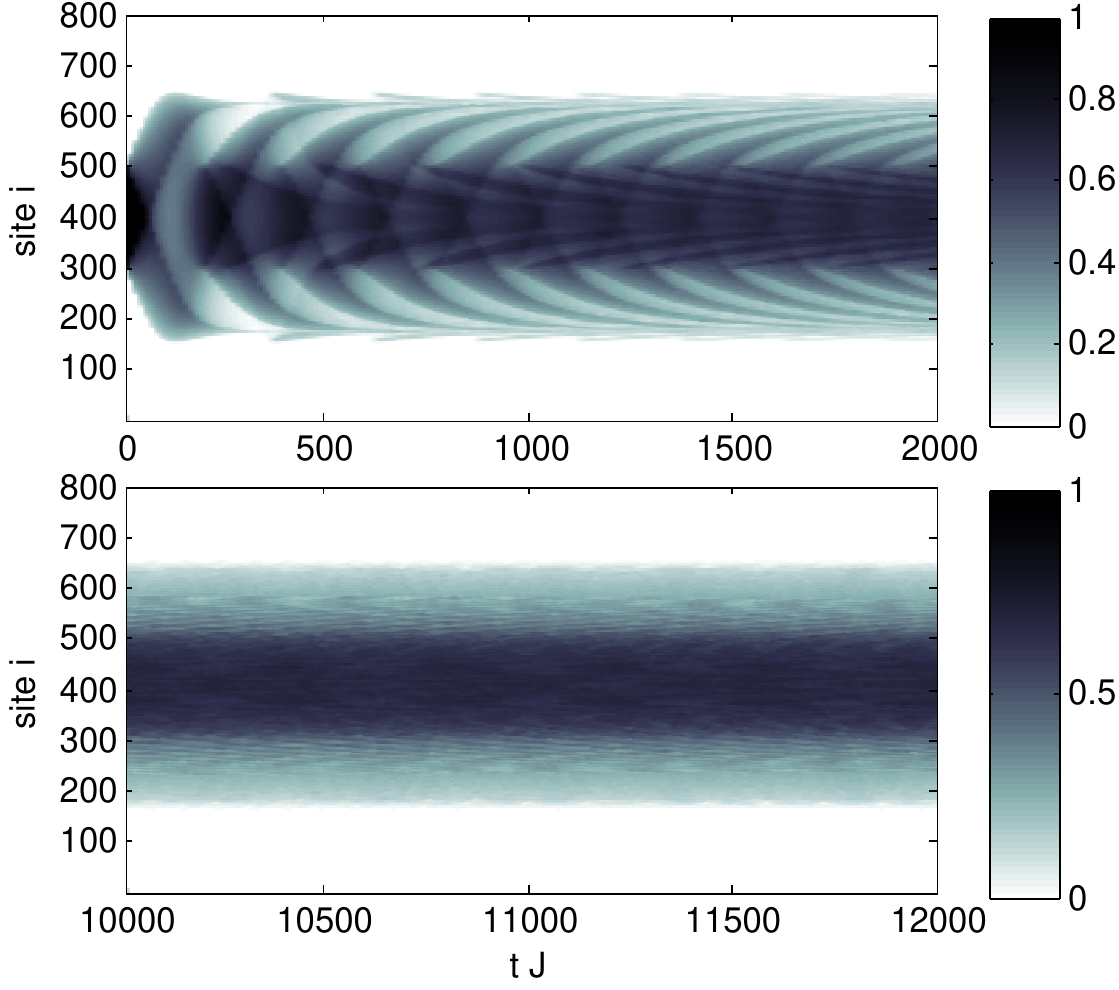}
\caption{~\CO Time evolution of the particle density for short times (upper plot) and long times (lower plot), showing $N=200$ bosons on $L=800$ lattice sites and $|V|=50/L^2$. 
Due to Eq. (\ref{eq:dens}), the dynamics of the particle density for $+V$ and $-V$ is the same and, apart from that, also identical for hard-core bosons and spinless fermions.
While the motion for short times appears to be coherent to a large extend, it becomes incoherent in the long time limit, where a quasi-stationary distribution emerges
(for a comparison with thermal reference ensembles, see Fig.~\ref{fig:mom_distr}). }
\label{fig:dens1}
\end{center}
\end{figure}

Before actually simulating this process numerically, let us address two issues. 
First, it is important to realize that the atomic cloud after the quench is still trapped, even by the inverted potential, $V<0$:
particles cannot get rid of their excess potential energy when trying to escape from the trap due to the boundedness of the kinetic energy~ \cite{negT}. 
Second, the sign change in $V$ and $U$ lifts the boundedness of the energy spectrum from below. 
Thus, {\it if} equilibrium is reached after the quench, it can only be at negative temperature~\cite{carr}. 

From those considerations alone, the actual relaxation process towards negative temperatures might still seem obscure. 
It turns out, however, that a symmetry of the Hubbard model helps to gain a better intuition. 
To this end, we will build on earlier works~\cite{expansion}, where this dynamic symmetry has been formulated for the homogeneous Fermi Hubbard model. 
For the homogeneous system ($V=0$), a \piboost operator $B$ was defined in Ref.~\onlinecite{expansion} which acts on the creation operators in lattice-momentum 
representation, $ B \, b_{\bf k} B^\dagger := b_{{\bf k} + {\bf Q}}$, where ${\bf Q} = (\pi,\ldots,\pi)$ is a $d$-dimensional vector (we set the lattice constant to one). 
This implies $B^2 = {\bf 1}$ and $B^\dagger = B$.
It follows, e.g. from Fourier transformation, that $B$ is strictly local in position space,
\be
B b_{\bf i} B^\dagger = \,  e^{i \,{\bf Q}\cdot {\bf i}} \, b_{\bf i} \,=\, (-1)^{i_1 + \cdots + i_d}\, b_{\bf i}
\ee
where ${\bf i} = (i_1,\ldots,i_d)$ is a lattice-site index. 
As a consequence, the boost operator leaves all local (on-site) terms invariant. 
Applying the boost operators to the inhomogeneous Hubbard model~(\ref{Hubbard}) thus only affects the sign of the kinetic term,
$B \cdot H(J,U,V) \cdot B^\dagger = H(-J,U,V)$, which can also be expressed as
\be
H(J,U,V) = - B \cdot H(J,-U,-V)\cdot B^\dagger  \label{eq:BHB}
\ee
Thus, a global sign change of the Hubbard model can be achieved by applying the $\pi$-boost operator and simultaneously inverting
the interaction strength and the trapping potential.
Now, let $\rho_0$ be the density matrix of the system at time $t=0$. 
Starting with $J=0$, the system is initially in a product state. Hence, the density matrix satisfies
\be
\rho_0 = B \rho_0 B^\dagger.
\label{eq:rho}
\ee
Now, let ${\mathcal O}$ be any time-reversal invariant observable.
Due to Eq.~(\ref{eq:BHB}) and the invariance of ${\mathcal O}$ and $\rho_0$ under time-reversal,
a combined action of time reversal, $\pi$-boost, and sign change of $U$ and $V$ leaves the
time evolution invariant:
\be
\langle {\mathcal O} \rangle_{H(J,U,V)}(t) = \langle B {\mathcal O} B^\dagger \rangle_{H(J,-U,-V)}(t) 
\label{eq:dynamic}
\ee 
A formal proof for the case $V=0$ is given in Ref.~\onlinecite{expansion}; its generalization
to arbitrary $V$ is a straightforward consequence of Eq.~(\ref{eq:BHB}).
Note that the $U\rightarrow \infty$ limit of hard-core bosons is included in the above more general statement.

Let us now focus on the one-dimensional case, relevant for our numerical studies, and let us 
address the most relevant operators in this context. For the particle density $n_i(t) = \langle b^\dagger_i b_i\rangle(t)$, 
the momentum distribution function $ n_k(t) = \frac{1}{N}\sum_{n,m}e^{i 2\pi (n-m)k/N} \langle b^\dagger_n b_m \rangle(t) $, 
and the total kinetic energy $e_{\rm kin}(t) = -J \sum_i \langle b_i^\dagger b_{i+1} + {\rm H.c.}\rangle$, the following symmetry conditions
follow from Eq.~(\ref{eq:dynamic}):
\be
n_i(t)_{H(J,U,V)} & = & n_i(t)_{H(J,-U,-V)} \label{eq:dens}\\
n_k(t)_{H(J,U,V)} & = &n_{k+\pi}(t)_{H(J,-U,-V)} \label{eq:nk}\\
e_{\rm kin}(t)_{H(J,U,V)} & = & - e_{\rm kin}(t)_{H(J,-U,-V)} \label{eq:ekin}
\ee
Hence, no difference between $V\rightarrow \pm V$ is observable in the time evolution for the particle density,
while the two momentum distributions evolve the same way up to a $\pi$-shift, and the kinetic energies have opposite signs for all times. 
As a consequence, given that the dynamics under $H(J,U,V)$ leads to a state at $T>0$ (characterized by a negative kinetic energy and
a momentum distribution peaked around zero), the dynamics under $H(J,-U,-V)$ then leads to a state with a momentum distribution 
peaked around $\pi$ and with a positive kinetic energy, i.e. a negative temperature state at $T<0$.

\section{Hard-core Bosons}
\label{sec:hcb}

\begin{figure}
 \begin{center}
  \includegraphics[width=\linewidth]{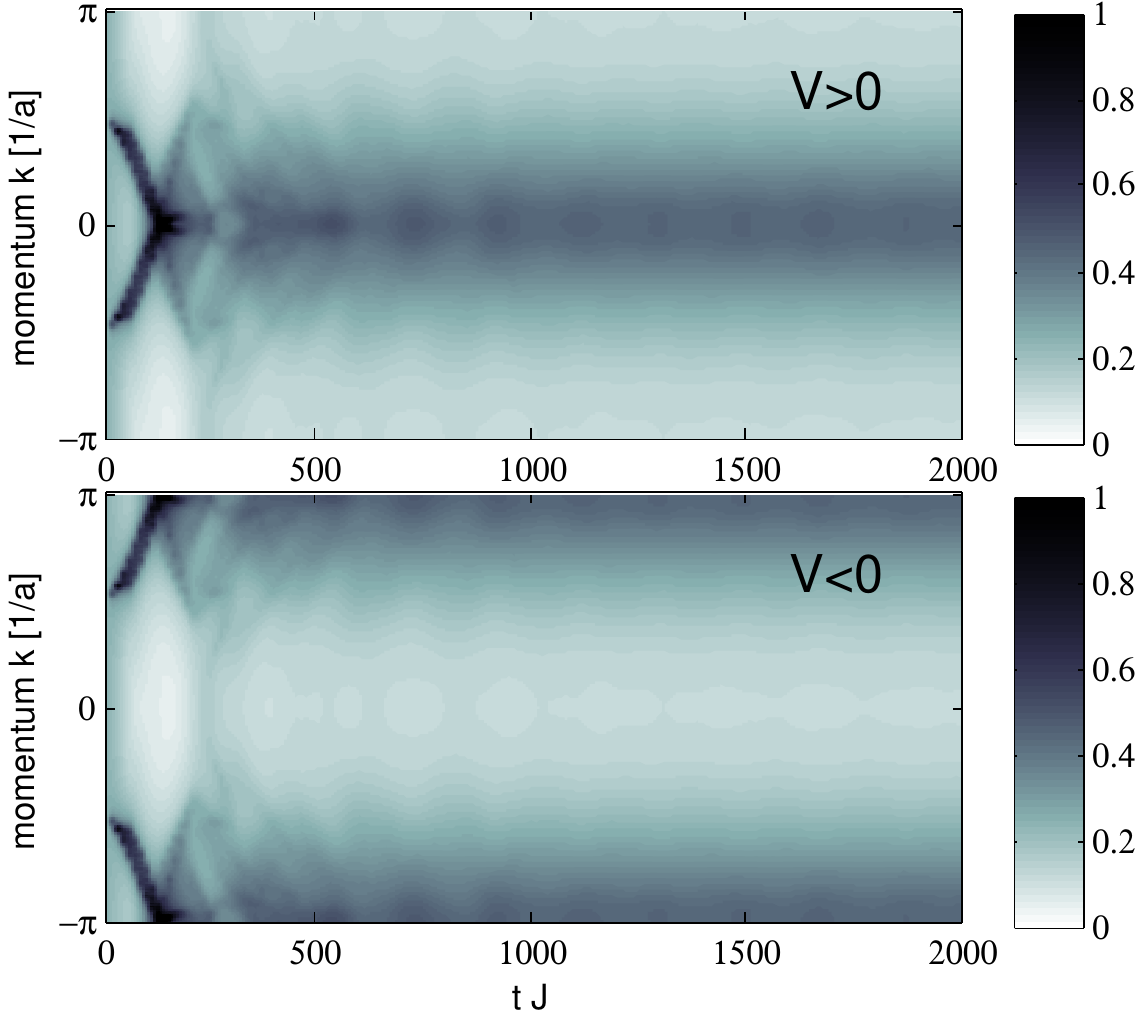}
  \caption{~\CO Momentum distribution functions as defined in Sec.~\ref{sec:model} after the quench from $J=0$ to $J=1$ and $V\rightarrow V$ (upper plot) and  $V\rightarrow -V$ (lower plot) for $N=200$ hard-core bosons
	  on $L=800$ lattice sites and $|V| = 50/L^2$. Starting from a constant momentum distribution in the initial state ('infinite temperature'),
	  a sharp peak around momentum zero or $\pi$, respectively, builds up on a short timescale which broadens at longer times. 
	  An enhancement of momentum states centered around zero ($\pi$) indicates positive (negative) absolute temperatures in the long time limit (see Fig.~\ref{fig:mom_distr} for the a comparison with equilibrium ensembles). %
	  } 
    \label{fig:mom}
\end{center}
\end{figure}

In order to be able to simulate large, spatially inhomogeneous systems for long times, we give our main attention to the $U\rightarrow \infty$ limit of the Hubbard model~(\ref{Hubbard}). 
Therefore, for the remainder of this paper with exception of Sec.~\ref{sec:ED}, the parameter $U$ will be omitted, and we will consider quenches
$V\rightarrow \pm V$ instead of $(U,V)\rightarrow (\pm U,\pm V)$. 

In one dimension, the system can be mapped to noninteracting spinless fermions,
and hence the problem can be formulated in terms of fermionic Slater determinants and Jordan-Wigner strings.
Pioneered by Rigol~\cite{rigolReview,rigol}, the method allows to study the time evolution for many lattice sites $L$ and large particle numbers $N$
with polynomial computational effort and for arbitrarily long times. 

To briefly summarize the numerical approach, let
$P^I$ be an $L \times N$ matrix, where the $N$ column vectors are the lowest $N$ fermionic eigenstates in position representation before the quench, thus characterizing our initial state.
Setting $\hbar=1$, the time evolution of the fermionic system after the quench is encoded in
$P(t) = U e^{i {\bf E} t}U^\dagger P^I$, where ${\bf E} = {\rm diag}(E_1,\ldots,E_n)$ is the diagonal matrix of single-particle energies {\it after} the quench, and U is the matrix that diagonalizes the 
fermionic Hamiltonian. As shown by Rigol e.g. in~\cite{rigolReview}, the bosonic single-particle density matrix $\rho_{ij} = \langle b_i^\dagger b_j\rangle$ can be calculated 
from $\{P_{ij}(t)\}$ in a series of manipulations that rely on the Jordan-Wigner transformation (summarized here from Ref.~\onlinecite{rigolReview}):
\be
\rho_{\alpha \beta}(t) & = & G_{\alpha \beta}(t) + \delta_{\alpha \beta}(1-2G_{\alpha \beta}(t)), \\
G_{\alpha\beta}(t) & = & \det[({\bf P}^\alpha)^\dagger(t) {\bf P}^\beta(t)], \n
({\bf P}^\gamma)_{ij} & = & \begin{cases}
                     -P_{ij} \; {\rm for} \;i<\gamma, j=1,\ldots,N \\
		      P_{ij} \;\;\; {\rm for}\; i\geq \gamma, j=1,\ldots,N \\
		      \delta_{i\gamma} \;\;\;{\rm for}\; j = N+1
                    \end{cases}\nonumber
\ee
where $\gamma = \alpha,\beta$ is a spatial index and ${\bf P}^\gamma$ an $L\times (N+1)$ matrix. Given $\rho_{ij}(t)$, we can calculate the time evolution of various bosonic observables.

We simulate the quench according to the protocol of Sec.~\ref{sec:model}, and we compare it to the case where we only quench
the hopping rate, but don't flip the trapping potential, i.e. we compare the cases $V\rightarrow \pm V$. 
According to Eq.~(\ref{eq:dens}), we expect identical time evolutions of the density profiles in both scenarios.
Figure~\ref{fig:dens1} shows the time evolution of the particle density for short and long times, respectively, which
by symmetry are identical for both quenches $V\rightarrow \pm V$ (not shown). In contrast, the time evolutions of the momentum distributions
agree upon shifting all momenta by $\pi$ in agreement with Eq.~(\ref{eq:nk}), see Fig.~\ref{fig:mom}.

\begin{figure}
 \begin{center}
  \includegraphics[width=\linewidth]{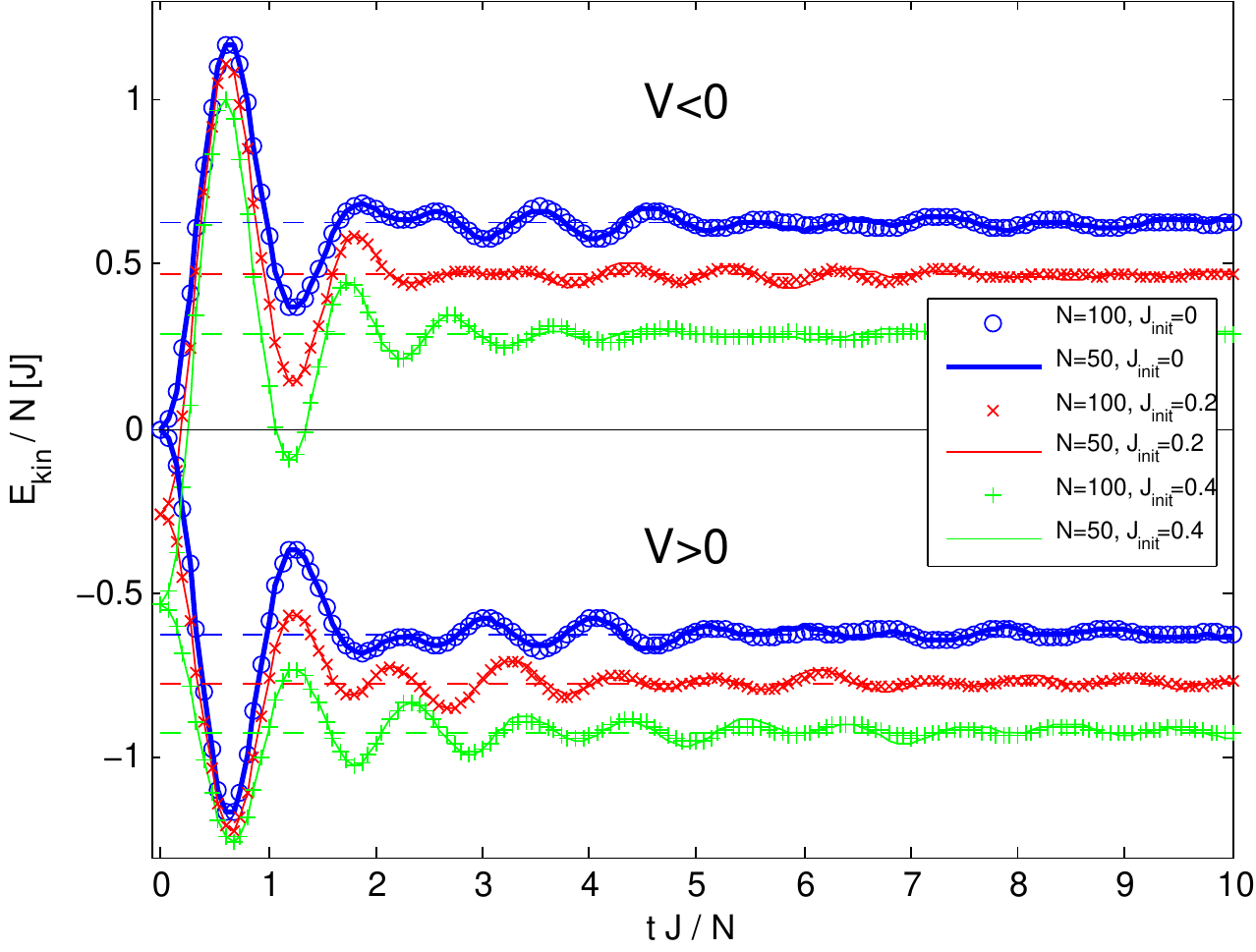}

  \caption{~\CO Dynamic symmetry and its breakdown: scaling plot of the total kinetic energy per particle as a function of time over particle number for different initial conditions:
	  $J_{init}=0$ (blue circles), $J_{init}=0.2$ (red x symbols) and $J_{init}=0.4$ (green ``+'' symbols). The lower curves correspond to the quench to $J=1,V\rightarrow V$ and show approximately equilibration to a state which can be associated to positive temperatures in the GGE in the sense discussed in the text (dashed lines: reference values independently obtained from the GGE).
	  The upper curves correspond to $J=1,V\rightarrow -V$ and (approximately) equilibrate in the same sense to negative temperature GGEs. Different system sizes $N=50,100$ are shown,
	  where the compression $(4N)^2|V|=50$ is kept constant.  At short times, the data collapse upon rescaling to a surprising degree.
	  } 
  \label{fig:ekin}
\end{center}

\end{figure}

Our approach can also be used to explore numerically the breakdown of the dynamic symmetry.
Of particular relevance is the violation of the symmetry requirement given in Eq.~(\ref{eq:rho}),
i.e., the case where the initial state is not a pure product state in position space, but, e.g., given by the system's ground state in the presence of
a small but finite $J_{init}>0$. In experiments, the hopping rate $J$ is controlled by the laser intensity, but it is not possible
to completely suppress the tunneling between neighboring lattice-sites.
Still, also in this case, the quenches from $J_{init}\neq 0$ to $J=1$ combined with $V\rightarrow \pm V$ will lead to 
positive and negative temperatures in the final states, respectively (say $T_1>0$ and $T_2<0$). However, $|T_1|$ will generally differ from $|T_2|$, and so do the 
time evolutions of various system observables for the two quenches.

Figure~\ref{fig:ekin} shows the time evolution of the kinetic energies per particle for various initial hopping rates $J_{init}\geq 0$ and for both quenches $V\rightarrow \pm V$.
In accordance with Eq.~(\ref{eq:ekin}), the case $J_{init}=0$ (blue curves) displays symmetric behavior 
for the two quenches: the time evolution of the total kinetic energies are the same up to a different sign.
In contrast, when starting from $J_{init} = 0.2$ (red curves) or $J_{init} = 0.4$  (green curves),
the symmetry condition in Eq.~(\ref{eq:rho}) is violated and the dynamic symmetry is broken. 
The figure also demonstrates that the
largest kinetic energies are reached for $J_{init}=0$. In equilibrium, large positive kinetic energies imply ``low'' negative temperatures, i.e. negative temperatures at low entropies.
According to Fig.~\ref{fig:ekin}, the lowest negative temperatures are reached for $J_{init}=0$. 
The figure also compares the final kinetic energies with the kinetic energies of reference equilibrium ensembles, which are discussed in the next section.

Note that Fig.~\ref{fig:ekin} also displays scaling properties of the hard-core bosons upon approaching the thermodynamic limit in the presence of the trap, i.e. $N\rightarrow \infty$, $V\rightarrow 0$, while $N^2V={\rm const}$.  
We compare $L=200$ with $L=400$ lattice sites, filling $N=L/4$ while keeping $|V|=50/L^2$ fixed. 
To a very high accuracy, at short times, the curves for different system sizes collapse 
as shown in Fig.~\ref{fig:ekin}, where time is scaled in units of the particle number $N$. However,
for longer times, when the kinetic energies become approximately stationary and only fluctuate around their mean values, 
the curves do not show this scaling behavior any more. Those fluctuations depend on the system size and get smaller for larger systems.

\section{Generalized Gibbs ensembles at negative temperature}

\label{sec:GGE}
\begin{figure}
 \begin{center}
  \includegraphics[width=\linewidth]{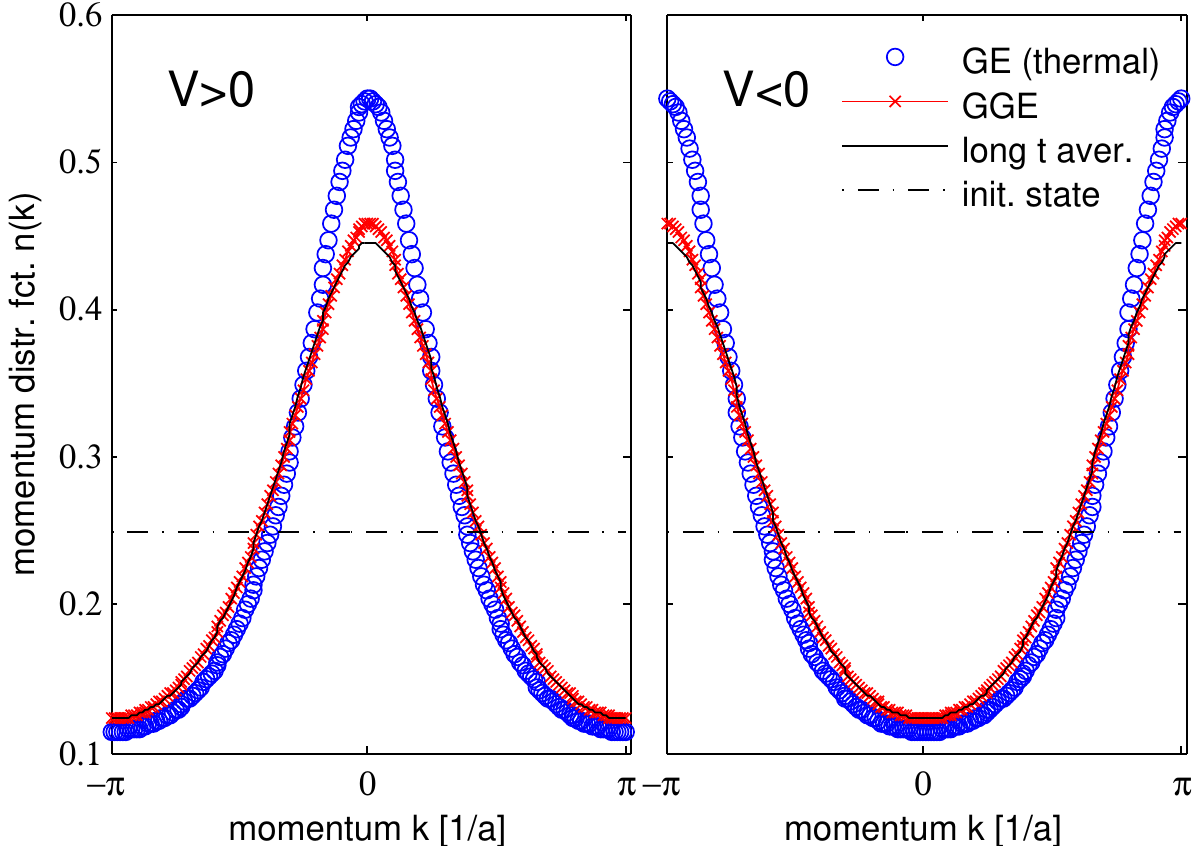}

  \caption{~\CO Momentum distributions of $N=50$ hard-core bosons on $L=200$ lattice sites in the final state after the quench $J=0 \rightarrow J=1$ and $V\rightarrow V>0$ (left) or $V\rightarrow V<0$ (right), $|V|=50/L^2$.
	  We compare the momentum distribution of the final state, averaged over times $1000 J^{-1}<t<2000 J^{-1}$ (black solid curve) with the Gibbs ensemble (``GE'', blue circles), where $\beta J=\pm 0.85 $ and $\mu = \pm 0.80 J$ are calculated
	  from energy and particle number conservation. We also compare the distributions to the generalized Gibbs ensemble (``GGE'', red x symbols). 
	  As predicted by the dynamic symmetry, the final momentum distributions for $V\rightarrow \pm V$ coincide upon shifting all momenta by $\pi$.
	  The discrepancies between the GGE and the long time average have been previously observed in Ref.~\cite{cassidi} and were attributed to finite size effects.} 
  \label{fig:mom_distr}
\end{center}

\end{figure}

\begin{figure}
 \begin{center}
  \includegraphics[width=\linewidth]{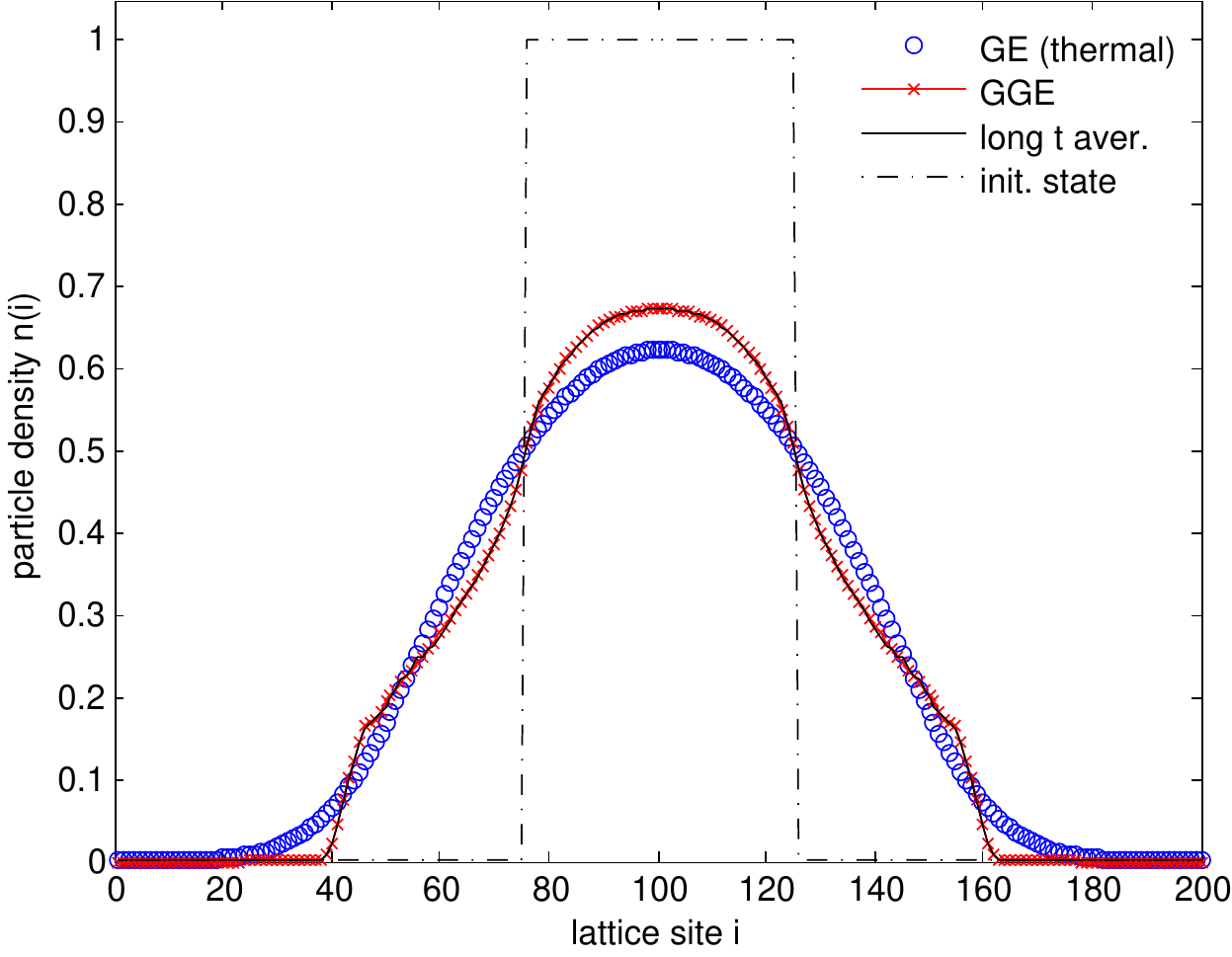}
  \caption{~\CO Density profile of the long time limit after the quench from $J=0$ to $J=1$ and $V\rightarrow \pm V$, showing the same parameters and ensembles as in Fig.~\ref{fig:mom_distr}.
	  The particle densities for $\pm V$ have been simulated and found to be identical as expected from the dynamic symmetry. The long time average (black solid line) coincides with the GGE prediction (red x symbols) to great accuracy, 
	  while deviations from the thermal Gibbs ensemble (blue circles) are pronounced. } 
 \label{fig:dens_equil}
\end{center}
\end{figure}

\begin{figure}
 \begin{center}
  \includegraphics[width=\linewidth]{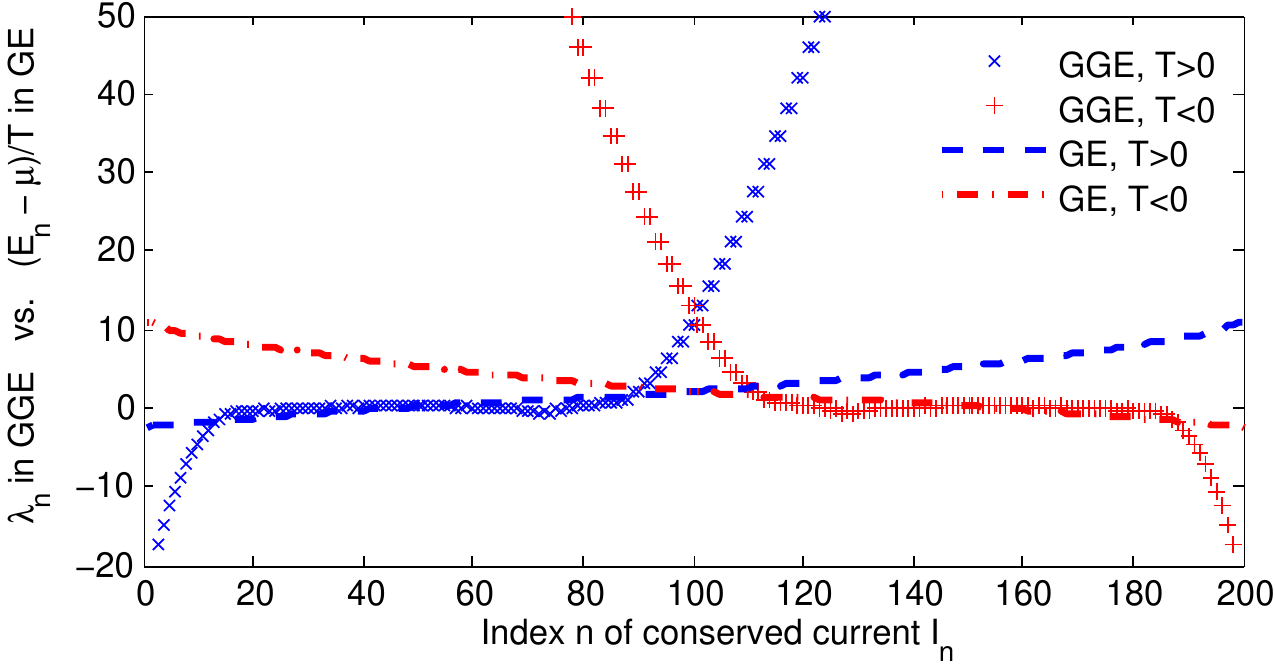}
 
  \caption{~\CO Lagrange multipliers $\lambda_n$ as a function of their index $n$ of the the generalized Gibbs ensemble (GGE).
	  The blue and rising (red and sloping) curves correspond to positive (negative) temperature states,
	  reached after the quench $J:0\rightarrow 1$ and $V\rightarrow V$ ($V\rightarrow -V$).
	  In order to compare the system to the Gibbs ensemble (GE) we also plot $(E_n - \mu)/T$ as a function of $n$.
	  The vector $(\vec{E} - \mu{\bf 1})/T$ can be understood
	  as a projection of $\vec{\lambda}$ on the subspace spanned by $\vec{E}$ and ${\bf 1}$, as discussed in the main text.
	  } 
\label{fig:Lagrange}
\end{center}
 
\end{figure}

After discussing the dynamic properties, let us now address the final state after the quench.
There is good evidence that certain integrable quantum systems, such as one-dimensional hard-core bosons, relax after a 
quench to a state described by a generalized Gibbs ensemble (GGE),
\be
\rho_{GGE} = \frac{1}{Z}e^{-\sum_n \lambda_n \hat{I}_n},
\ee
see Ref.~\onlinecite{rigolGGE} for more details.
Here, the $\hat{I}_n$ are the conserved quantities of the system and $\lambda_n$ the corresponding Lagrange multipliers.
As one-dimensional hard-core bosons can be mapped to noninteracting fermions, the operators $\hat{I}_n$ have a simple meaning in 
their fermionic representation: they are the projectors onto the fermionic single-particle eigenstates at energies $E_n$, 
i.e. $\hat{I}_n = c_n^\dagger c_n$ where $c^\dagger_n, c_n$ are the fermionic creation and annihilation operators of the $n_{th}$ energy eigenstate, respectively.
Their expectation values $\langle \hat{I}_n\rangle$ do not change in the course of time.
The Lagrange multipliers are obtained from $ \lambda_n = \ln \left[(1-\langle \hat{I}_n \rangle)/\langle \hat{I}_n \rangle\right] $,
see, e.g., Ref.~\onlinecite{rigolReview}.  
In contrast, a thermal Gibbs ensemble (GE) is characterized only by the two Lagrange multipliers chemical potential $\mu$ and (inverse) temperature $\beta=1/T$. 
When calculating the temperature for a given particle number $N$ and energy $E$ for the hard-core bosons in thermal equilibrium, the Bose-Fermi mapping can be used.
Here, chemical potential $\mu$ and inverse temperature $\beta$ are determined by
\be
N & = & \sum_n \frac{1}{1+\exp(\beta (E_n -\mu))} \\
E & = & \sum_n \frac{E_n}{1+\exp(\beta (E_n -\mu))} \nonumber
\ee
where $E_n$ are the fermionic single-particle energies. 
The equations are then solved for $\mu$ and $\beta$ by using, e.g. Newton's algorithm \cite{NumRec}.

Let us now compare the GE to the GGE and introduce negative temperatures in the context of the GGE.
We know that the noninteracting, many-body fermionic Hamiltonian can be expressed as $H=\sum_n E_n \hat{I}_n$, 
and as the $\hat{I}_n$ are projectors to eigenstates, we also have $N = \sum_n \hat{I}_n$. 
Hence, we can write the Gibbs ensemble in analogy to the GGE as
\be
\rho_{GE} = \frac{1}{Z}e^{-\beta(H-\mu N)} = \frac{1}{Z}e^{- \sum_n \beta (E_n - \mu)\hat{I}_n} \label{eq:GE}.
\ee

Note that the notion of temperature can also be defined uniquely for the GGE, namely by adjusting $\mu$, $\beta$ and $\vec{\delta}$ for a given $\vec{\lambda}$ such that
\be
\beta(\vec{E} - \mu{\bf 1} + \vec{\delta}) \stackrel{!}{=} \vec{\lambda}  \label{eq:stackrel}
\ee
where ${\bf 1} = (1,\ldots,1)^t$ and $\vec{\delta}$ is an $L$-dimensional vector in an $(L-2)$-dimensional subspace, satisfying $\vec{\delta} \perp {\bf 1}$ and $\vec{\delta} \perp \vec{E}$.
For finite $L$, we can solve this equation for $\mu$ and $\beta$ by projecting Eq.~(\ref{eq:stackrel}) onto the two subspaces spanned by ${\bf 1}$ and $(\vec{E}-\mu {\bf 1})$, 
\be
\mu_{\rm GGE} & = & \frac{1}{L}[(\vec{E}\cdot {\bf 1}) - \beta^{-1} (\vec{\lambda}\cdot {\bf 1})] \label{eq:defT} \\
\beta_{\rm GGE} & = & \vec{\lambda}\cdot (\vec{E}-\mu{\bf 1})/[(\vec{E}-\mu{\bf 1})\cdot (\vec{E}-\mu{\bf 1})]  \nonumber
\ee
where we used the orthogonality of $\vec{\delta}$ to those subspaces and $L = ({\bf 1}\cdot {\bf 1})$. 
The concept of assigning a temperature to a generalized Gibbs ensemble is not very deep, as temperature
is only one of $L$ Lagrange multipliers that characterize the system. However, its definition by Eq.~(\ref{eq:defT})
allows for an intuitive picture: the inverse temperature
$\beta_{\rm GGE}$ expresses the degree of alignment or anti-alignment of the vector of Lagrange multipliers with the vector of energies,
measured relative to the chemical potential, see also Fig.~\ref{fig:Lagrange}.
We can therefore speak of generalized Gibbs ensembles at negative absolute temperatures.

To calculate the GE and the GGE numerically, we use the following expressions derived in Ref.~\onlinecite{rigolFiniteT}:
\be
\rho_{ij} & = & \frac{1}{Z}\left(\det\left[{\bf 1} + ({\bf 1}+A)O_1 U D U^\dagger O_2\right]\right.\\
	  &  & - \left. \det \left[ {\bf 1} + O_1 U D U^\dagger O_2\right]\right), \quad i\neq j \n
\rho_{ii} &=& \left[U(1 + D)^{-1}U^\dagger\right]_{ii}	  
\ee
Here, $A_{ij}=1$ and elsewhere contains only zeros, and $O_1(O_2)$ is diagonal with the first $j-1 (i-1)$ elements being $-1$ and all others $+1$ respectively.
Furthermore, the matrices $U,U^\dagger$ diagonalize the single-particle Hamiltonian, and $D$ is a diagonal matrix that depends on the ensemble under consideration [$D = e^{-\beta(\vec{E}-\mu{\bf 1})}$ for the GE and $D = e^{-\vec{\lambda}}$ for the GGE]. 

Our numerical simulations give strong evidence that after the quench, the physical system relaxes to a generalized Gibbs ensemble, see Figs.~\ref{fig:dens_equil} and \ref{fig:mom_distr}. 
Figure~\ref{fig:dens_equil} shows the particle density profiles in the long time limit, being identical for $\pm V$. 
As expected due to the system's many conservation laws, the long time averaged density profile (black curve) does not look like a thermal distribution (blue), 
but instead very closely matches the GGE prediction (red curve). 
Figure~\ref{fig:mom_distr} shows the momentum distribution after the quench from $J=0$ to $J=1$ and $V\rightarrow V$ (left) and $V\rightarrow -V$ (right). 
In both cases, discrepancies between the long time average and the thermal ensembles are visible.
Up to small deviations, the long time averaged data are instead very close to the GGE prediction, involving positive temperatures (left) and negative temperatures (right).
Similar small deviations have been observed in~\cite{cassidi}, which were attributed to finite size effects.

It is also interesting to compare the Lagrange multipliers of the generalized Gibbs ensembles for both quenches, i.e. $V\rightarrow \pm V$, shown in Fig.~\ref{fig:Lagrange}. 
To also allow for a comparison with the thermal ensembles, we plot the $L$ dimensional vector $\beta(\vec{E}-\mu {\bf 1})$, that is the closest related quantity for the GE: due to Eq.~(\ref{eq:stackrel}), 
this vector can be understood as the projection of $\vec{\lambda}$ to the subspace spanned by ${\bf 1}$ and $\vec{E}$, thus denoting the reference ensemble with the same particle number and total energy. 
The plot shows that the corresponding curves for positive and negative temperatures map onto each other when inverting the labeling of the index $n$.

Figure~\ref{fig:Lagrange} reveals striking deviations between the GE and the GGE for very large and very small indices $n$.
Explaining those deviations, we focus on $T>0$ (the discussion of $T<0$ is analogous). 
It is important to keep in mind that the Lagrange multipliers are determined by the overlap of the initial state with the eigenstates
of the final Hamiltonian. As our initial state is tightly centered around the origin of the lattice, it has a very large overlap with 
low energy states (at small $n$), which are also centered around the origin. This explains that the corresponding Lagrange multipliers are large and negative, 
implying an enhanced occupation of those states compared to the thermal ensemble, which allows for energy exchange.  For large $n$ (at $T>0$), the Lagrange multipliers come in pairs, reflecting the Bloch-localized states at the right and 
left edge of the harmonic trap, see, e.g.,~\cite{rigol:bloch, rey:bloch}. As our initial state has vanishing overlap with those states, the corresponding Lagrange multipliers are large and positive.

\section{Exact Diagonalization}
 \label{sec:ED}

\begin{figure}
 \begin{center}
  \includegraphics[width=\linewidth]{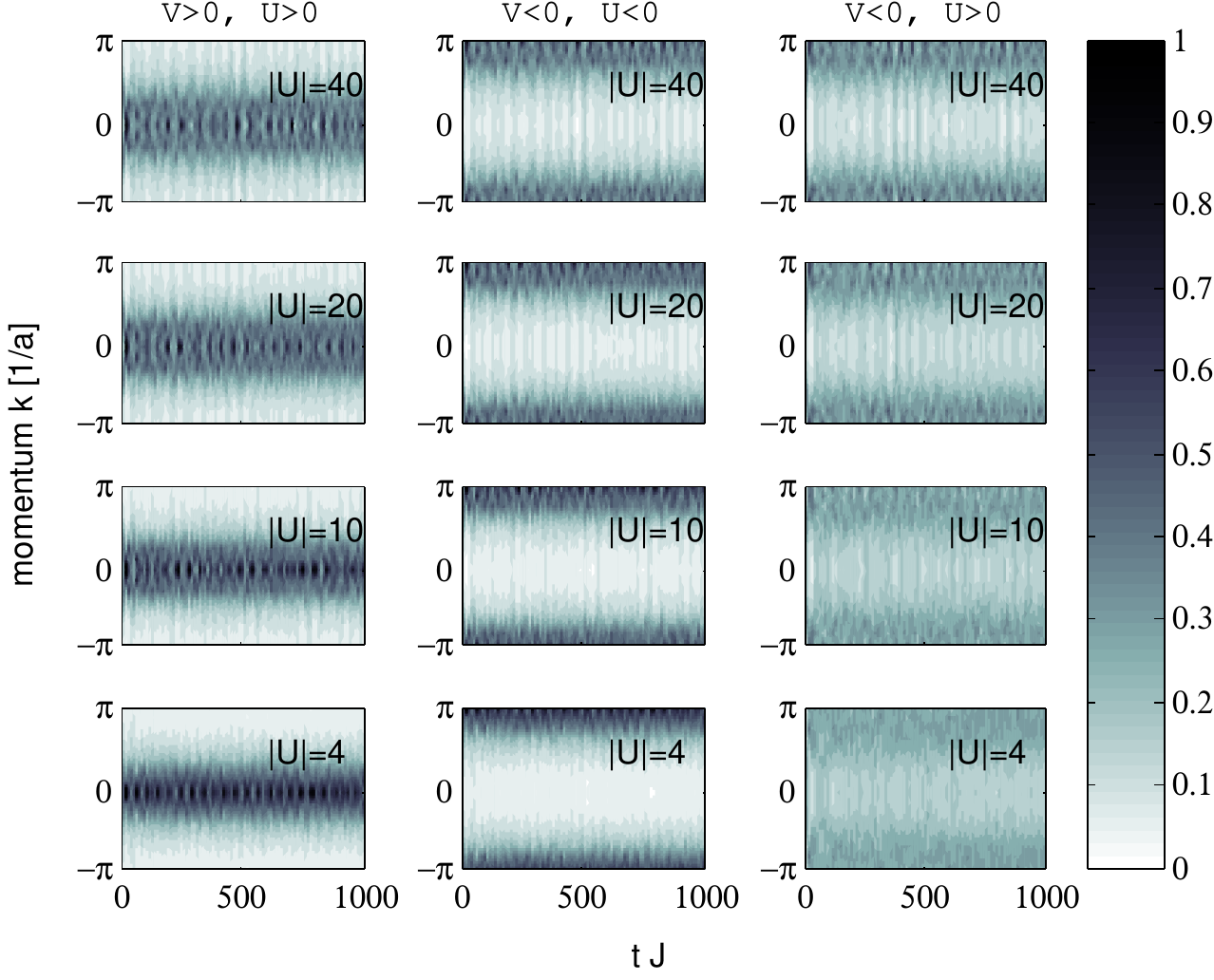}
\caption{~\CO Momentum distribution functions for $N=4$ particles on $L=16$ lattice sites (trapping potential $|V|=50/L^2$).
	The rows correspond to various initial interaction strengths $U/J = 40, 20, 10, 4$, whereas the columns correspond to the quenches
	$J:0\rightarrow 1$ combined with quenches from 	$(U,V)$ to $(U,V)$ (left), $(-U,-V)$ (middle) and $(U,-V)$ (right). While the first two quenches
	are characterized by peaks around zero and $\pi$, the latter case is expected to be unstable and shows a 'washed-out' momentum distribution. Note that when $|U|$ is much larger than all other energy scales, 
	differences between the last two columns become smaller, i.e. both quenches approach the hard-core limit for $|U|\gg J$.}
\label{fig:ED}
 \end{center}
\end{figure}
To complement our discussion of the integrable case, we also simulated the nonintegrable case of finite on-site interaction. 
To do so, we applied the adaptive time-dependent density matrix renormalization group (DMRG) \cite{white1992,schollwoeck2005,schollwoeck2011,Daley2004,White2004} 
and a time-dependent exact diagonalization (ED) approach relying on a Krylov-space approximation of the time evolution operator \cite{Park1986,Hochbruck1997,Moler2003,Manmana2005}. 
The kind of quenches studied in this work represent the worst case scenario for a time-dependent DMRG simulation \cite{Hamma}.
Due to the dramatic entanglement growth, it is not possible to reach long times beyond $\sim 10-20/J$ using the DMRG. 
Therefore, we focus here on the long-time behavior of results obtained using the ED with four particles on 16 lattice sites and present them as a proof-of-principle calculation. 

Fig.~\ref{fig:ED} shows the time evolution of the momentum distributions for different interaction strengths and for the three quenches $J:0\rightarrow 1$, 
$(U,V) \rightarrow (U,V),(-U,-V)$ and $(U,-V)$. 
The latter case can not be expected to relax to a thermal state, as the underlying Hamiltonian is unbounded. 
In contrast, in the two other cases, the nonintegrable 1D Bose-Hubbard model at finite $U$ shows relaxation to a 
quasi-stationary momentum distribution that shares the expected features of negative temperatures discussed for the integrable limit, 
i.e. a momentum distribution peaked around $\pi$ for $U<0,V<0$. 

A detailed analysis of the instabilities of the system occurring for $U>0,V<0$ (right column of Fig.~\ref{fig:ED}) are an interesting topic in their own right, 
which we leave open for future studies. 
Note, however, that if $U$ is much larger than all other accessible energy scales, double-occupancies are highly suppressed and
even the unstable system shows qualitatively similar behavior to the stable system at $U<0$,$V<0$ as can be seen by comparing 
the second and third picture in the first row of Fig.~\ref{fig:ED} (showing $|U|/J=40$).

\section{Summary}
\label{sec:summary}

Our results indicate that
hard-core bosons in a one-dimensional optical lattice and trapped by a harmonic potential equilibrate to a generalized Gibbs ensemble
at negative absolute temperature after turning the trap $V$ into an anti-trap, $V\rightarrow -V$. 
Inverse temperature is only one of many Lagrange-multipliers that characterize the integrable system, but can be explicitly defined (see Eq.~(\ref{eq:defT})): 
it describes the degree of (anti-) alignment of the vector of Lagrange multipliers to the vector of eigen-energies, measured relative to the chemical potential
(another Lagrange multiplier). States at negative temperature in the generalized Gibbs ensemble show qualitatively similar features as in the Gibbs ensemble: 
both are characterized by positive kinetic energies and have momentum distributions that are peaked around the momentum $\pi$ (rather than $0$), showing
an enhanced occupation of the band maxima (note that the latter was used in Ref.~\cite{braun} as an identifier of negative temperatures in a Bose-gas).
Hence, the experimental detection of negative-temperature hard-core bosons should be no more difficult than in the nonintegrable case. From a theoretical perspective, 
the integrable limit allowed us to study the full quantum dynamics of a large system. \\
In the first part of this paper, we aimed at shedding light on the equilibration process towards negative absolute temperature states in the Bose-Hubbard model 
in any dimension and any value of $U/J$. We showed that a dynamic symmetry helped to understand the equilibration process for a subclass of initial states such as product states,
relevant in the recent experimental realization of $T<0$~\cite{braun}. To this end, we compared the quench in the hopping rate $J:0\rightarrow 1$ and fixed $V>0$ to the
quench $J:0\rightarrow 1$ combined with $V\rightarrow -V$. Due to the dynamic symmetry,
the evolution of the kinetic energies of the two systems show the same behavior for all times, up to having opposite signs. 
The same occurs with the momentum distributions, up to a shift by $\pi$.
We illustrated the validity of this symmetry and its breakdown ($J_{init} \neq 0$) using our simulations for hard-core bosons.\\
In the future, it will be relevant to investigate stability aspects of negative temperatures when going away from the
idealized case of an isolated Hubbard model, where negative temperature states are perfectly stable just as positive temperature states. 
Here, we suggest to study the coupling to a thermal bath or to study the impact of higher bands of an optical lattice.
\section*{Acknowledgments}
We would like to thank D. Huse, M. Schiro, M. Rigol, A. Rapp, and A. Rosch for helpful discussions.
S.M. acknowledges financial support from the NSF MRSEC program through the Princeton Center for Complex Materials Fellowship (DMR-0819860), 
the ICAM travel award (DMR-0844115), and the DARPA OLE program. A.E.F. thanks NSF for support under Grant No. DMR-1339564.

\end{document}